\documentstyle[preprint,aps,eqsecnum,epsf]{revtex}
\draft
\def\Spicture#1#2#3{\begin{tabular}{c}{\epsfxsize=#1\epsfbox{#2}}\\
\mbox{#3}\end{tabular}}

\tightenlines

\begin{document}

\title{Practical approximation scheme for the pion dynamics \\
        in the three-nucleon system}
\author{L.~Canton$^{a,b}$, T.~Melde$^{b,c}$, and J.P.~Svenne$^{b,c}$}
\address{$^{a}$ Istituto Nazionale di Fisica Nucleare, 
                Padova I-35131, Italy \\
       $^{b}$ Physics Department, University of Manitoba, 
              Winnipeg, MB Canada\\ 
       $^{c}$ Winnipeg Institute for Theoretical Physics,
       Winnipeg,  MB Canada \\
       }
\date{April 10, 2000}

\maketitle

\begin{abstract}
We discuss a working approximation scheme to a recently
developed formulation of the coupled $\pi$NNN-NNN problem.
The approximation scheme is based on the physical assumption that,
at low energies, the 2N-subsystem dynamics in the elastic channel
is conveniently described by the usual 2N-potential approach, while
the explicit pion dynamics describes small, correction-type effects.
Using the standard separable-expansion method, we obtain a dynamical
equation of the Alt-Grassberger-Sandhas (AGS) type. This is an
important result, because the computational techniques used
for solving the normal AGS equation can also be used to describe the
pion dynamics in the 3N system once the matrix dimension is increased
by one component. We have also shown that this approximation scheme
treats the conventional 3N problem once the pion degrees of freedom
are projected out. Then the 3N system is described with an extended
AGS-type equation where the spin-off of the pion dynamics (beyond
the 2N potential) is taken into account in additional contributions
to the driving term. These new terms are shown to reproduce the
diagrams leading to modern 3N-force models. We also recover two
sets of irreducible diagrams that are commonly neglected in
3N-force discussions, and conclude that these sets should be further
investigated, because a claimed cancellation is questionable.
\end{abstract}

  \hfill {Preprint DFPD 00/TH/22  }

\section{Introduction}

Considerable progress has been made in understanding the coupled
system of two nucleons and (at least) one pion. This enables the study
of  pion absorption and production processes on very light nuclei, and is a
next step of including explicit pion degrees of freedom, beyond
the standard nuclear picture where mesonic degrees of freedom are ``frozen
out'', into nucleon-nucleon potentials. We are in particular
interested in a set of theories which can be classified
by the acronym TRABAM (Thomas-Rinat, Afnan-
Blankleider, Avishai-Mizutani), Ref.~\cite{trabam}.
In recent years, a fair amount of effort has been made
to extend this theory to the pion-three-nucleon
domain, where a richer range of phenomena is possible. In a sequence of
papers \cite{cancat}, this system has been explored in the attempt to
arrive at a consistent and connected theory of the coupled
$\pi$NNN-NNN system. This effort has required the blending
of the standard three- and four-body theories of Alt, Grassberger
and Sandhas (AGS) \cite{ags},
    with the possibility that a pion can appear and disappear
anywhere in the system, and in any of its subsystems.

In a recent paper, one of the authors \cite{Canton98} elucidated a connected,
coupled scheme for the combined $\pi$NNN-NNN dynamics.
Furthermore, he derived, by the use of the quasi-particle formalism at
three-cluster and two-cluster levels, an equation for the coupled
$\pi$NNN-NNN system, that has the appearance of a coupled set of
Lippmann-Schwinger type equations:
\begin{equation}
X_{ss'}^{(2)}=Z_{ss'}^{(2)}+\sum\limits_{s''}
{Z_{ss''}^{(2)}\tau_{s''}^{(2)}X_{s''s'}^{(2)}} \, . \label{eLe}
\end{equation}
An equation of this type was first given by Lovelace \cite{love} in
the standard
three-particle problem with separable interactions, using the original theory
by Faddeev \cite{fadd}.
It was also derived by AGS
\cite{ags} who applied the quasiparticle approximation to their
three-body equations. Equation (\ref{eLe}), like the
Faddeev-Lovelace-AGS equation,
is a coupled set of integral equations in one
inter-cluster momentum variable, but here the labels
$s\;(s',\;s'')$
     run over four values, rather than three:
$s=0$
     representing the configuration consisting of a three-nucleon cluster
with the pion separate,  and
$s=1,\,2,\,3$
     representing the three possible arrangements of the three nucleons in a
pair and a separated nucleon (the usual AGS-Lovelace scheme) with the
pion associated with either the pair or the single nucleon, or absent. (See
table II in \cite{Canton98}.)

The driving term in this equation
is given in terms of quantities associated with the two levels of
separable approximations needed when the quasiparticle method is applied
\cite{Canton98}
to the full $\pi$NNN-NNN system of equations, as follows
     \begin{eqnarray}
Z_{ss'}^{(2)}&=&\left\langle {\left( {s^{(2)}} \right)_-}
\right|g_0\left| {\left( {{s'}^{(2)}} \right)_-} \right\rangle \bar 
\delta _{ss'}
\label{driv} \\
      &+&\sum\limits_{a'(\subset s)} {\,\sum\limits_{b'(\subset s')}
{\,\sum\limits_{a(\subset a',b')} {\left\langle {\left( {s^{(2)}}
\right)_{a'a}} \right|\tau _a^{(3)}\left| {\left( {s'^{(2)}} \right)_{b'a}}
\right\rangle (}}}\bar \delta _{ss'}+\delta _{ss'}\bar \delta _{a'b'}).\cr
\nonumber
\end{eqnarray}

This equation looks rather complicated with the multiple sums and
inclusion rules, but, in fact, it turns out to be rather simple, once all these
rules are applied and the delta-functions invoked. We will define in detail
symbols in this equation below, but let us first look at the structure of this
driving term.
The first term contains only components from
the three-nucleon (no-pion) sector, and therefore cannot contribute at all
if  $s=0$ or $s'=0$. Therefore, the only contributions to this term
can come from the case  $s\ne 0$ and $s'\ne 0$. In this case the first term of
   Eq.~(\ref{driv}) does not contribute on the diagonal (i.e. $s=s'$)
because of the anti-delta function. Off the diagonal, since this contains
only nucleon degrees of freedom, this term can properly be identified with
the driving term of the traditional AGS-Lovelace approach to the standard
three-nucleon problem.

For $s=s'=0$ there is no contribution to $Z_{ss'}^{(2)}$
also from the second term,
because the inclusion prescriptions in the sums cannot be satisfied if
$s=s'=0$.
More explicitly, since
$s=s'$,
only the last term in the delta-function structure
survives, but this requires
$a'\ne b'$.
However, for
$s=0$,
there is only one two-cluster state, namely
$\pi \,(N_1\,N_2\,N_3)$,
so the condition
$a'\ne b'$
cannot be satisfied.
Therefore, the driving terms do not contribute at all in the case
$s=s'=0$, either in the first, or in the second term.

The second term does have non-zero values
in the diagonal case $s=s'\ne 0$, and this provides
additional diagonal terms to  Eq.~(\ref{eLe}) due
to the coupling to the pion degrees of freedom.
Even if small, such contributions
compare to zero and therefore can hardly be neglected.
The second term contributes also
to the off-diagonal elements, thus providing corrections
to the dominant terms of the standard three-nucleon problem.
%

In Eq. (\ref{driv}),
$\left| {\left( {s^{(2)}} \right)_{a'a}} \right\rangle ,\;\left| {\left(
{s^{(2)}} \right)_-} \right\rangle $
     are components of the form-factor vector coming from the separable
approximation at the two-cluster level, the subscript ``-'' representing the
no-pion sector, the subscripts
$a'a$
being the Yakubovsk{\u\i}\cite{yakub} chain labels for the chains of
partition of the (conserved) four-body system, $\pi$NNN. The quantity
$g_0$
is the free propagator for three nucleons, in the absence of pions; the
quantity
$\tau _a^{(3)}$
denotes the intermediate propagation of the possible
three-cluster structures of the $\pi$NNN system.
In other words, the separable representation
of the two-particle $t$-matrix,
according to
\begin{equation}
t_a(z)\cong \left| {\left( {a^{(3)}(z)} \right)} \right\rangle \tau
_a^{(3)}(z)\left\langle {\left( {a^{(3)}(z)} \right)} \right|
\label{apole} \, ,
\end{equation}
describes effectively all the elastic two-body processes in the
4-body space, with $t_a$ representing either the
$NN$ or the $\pi N$ two-body $t$-matrix.
The ``anti-delta'' function is:
$\bar \delta _{ss'}=1-\delta _{ss'}$.
The inclusions under the summation symbols are intended in the
usual sense of Yakubovsk{\u\i} chain inclusions. For example, for
$s=1,\;a'$
is one of the two partitions
$N_1\,(N_2\,N_3\,\pi )$, $(\pi \,N_1)\,(N_2\,N_3)$,
and $a$ are all the possible three-cluster partitions that can be obtained by
breaking one cluster in each of the above two-cluster partitions.
As an example of how the complicated-looking structure of Eq.~(\ref{driv})
simplifies, we show two particular contributions to
the driving term,
where we specify the form-factor vectors by the
chains of partition, and leave off the superscripts
(2) and (3) that occur in Eq.~(\ref{driv});
a typical off-diagonal element:
\begin{eqnarray*}
\ \ \ \ \ \ Z_{12}=&
\left\langle  N_1(N_2N_3) \left| g_0 \right| N_2(N_3N_1)\right\rangle
  &\hfill\hfill\hfill\ \ \ \ \ \ \ \ \ \  (1.2a) \\
+&\left\langle  N_1(N_2N_3\pi);N_1N_2(N_3\pi) \left|
\tau_{(N_3\pi)} \right| N_2(N_3N_1\pi);N_1N_2(N_3\pi)\right\rangle \\
+&\left\langle  N_1(N_2N_3\pi);N_1N_3(N_2\pi) \left|
\tau_{(N_2\pi)} \right| (\pi N_2)(N_3N_1);N_1N_3(N_2\pi)\right\rangle \\
+&\left\langle  (\pi N_1)(N_2N_3);N_2N_3(N_1\pi) \left|
\tau_{(N_1\pi)} \right| N_2(N_3N_1\pi);N_2N_3(N_1\pi)\right\rangle \, ,
\end{eqnarray*}
and a diagonal element:
\begin{eqnarray*}
\ \ \ \ \ \ \ Z_{11}=&
\left\langle {N_1\left( {N_2N_3\pi } \right);N_1\left( {N_2N_3}
\right)\pi } \right|\tau _{\left( {N_2N_3} \right)}\left| {\left( {N_1\pi }
\right)\left( {N_2N_3} \right);N_1\left( {N_2N_3} \right)\pi }
\right\rangle
&\hfill\ \ \ \ \ \ \ \ \ \ \ \ \ (1.2b) \\
+&\left\langle {\left( {N_1\pi } \right)\left( {N_2N_3} \right);N_1\left(
{N_2N_3} \right)\pi } \right|\tau _{\left( {N_2N_3} \right)}\left|
{N_1\left( {N_2N_3\pi } \right);N_1\left( {N_2N_3} \right)\pi }
\right\rangle   \, .
\end{eqnarray*}
For $s=1$,
as can be immediately deduced by comparing Eq.~(\ref{driv})
with Eq.~(1.2$a$), the state $\left|(s)_{-}\right\rangle$
denotes the Faddeev component $\left|N_1(N_2N_3)\right\rangle$, while with
$\left|(s)_{a'a}\right\rangle$ we denote the relevant Yakubovsk{\u\i}
components, such as
$\left|N_1(N_2N_3\pi);N_1N_2(N_3\pi)\right\rangle$ for instance.
All possible 4-body components for $s=1$ are listed in Tab.~\ref{a'a}.

To solve Eq.~(\ref{eLe}) one must first define and construct the states
$\left| {\left( {s^{(2)}} \right)_{a'a}} \right\rangle$, $\left| {\left(
{s^{(2)}} \right)_-} \right\rangle$, and the two-cluster Green's function
$\tau_s^{(2)}$. It is precisely at this point that we propose
herein a workable approximation scheme. However, before
discussing the approximation, we recall for clarity
the rigorous result obtained in Ref.~\cite{Canton98}.
For complete details we refer to that work.

Starting from the two-particle representation Eq.~(\ref{apole}),
one can derive the dynamical sub-amplitude containing the
interactions internal to the coupled set of partitions
$(N_1) (N_2 N_3)$, $(N_1\pi) (N_2 N_3)$, and $(N_1) (N_2 N_3\pi)$.
This is denoted by $({\bf x})_{(s=1)}$ where the $s=1$
index should be interpreted with the fact that nucleon ``1" is set
apart from the other two: obviously, the other cases with
$s=2,3$ are obtained by cyclic permutations of the nucleons.
The $s=0$ amplitude instead contains all the interactions
internal to the single $\pi (N_1 N_2 N_3)$ partition, hence the pion is
set apart from the three nucleons here.

The dynamical equations for these four subamplitudes are given by
(see Eqs.~(3.21-24) of Ref.~\cite{Canton98})
\begin{eqnarray}
\left( {x_s} \right)_{a'a,b'b}=&\langle a^{(3)} | G_0 | b^{(3)} \rangle
\bar \delta _{ab}\delta _{a'b'}
+\sum\limits_{c'(\subset s)} {\sum\limits_{c(\subset c')}
\langle a^{(3)} | G_0 | c^{(3)} \rangle
    {\bar \delta _{ac}\delta _{a'c'}\tau^{(3)}_c
\left( {x_s} \right)_{c'c,b'b}}} \nonumber \\
&+\langle a^{(3)}|G_0 \left( {f_s} \right)_{a'a}
g_0\left( {x_s}^\dagger \right)_{-,b'b} \label{first} \\
\left( {x_s}^\dagger \right)_{-,b'b}=&\left( {f_s^\dagger } \right)_{b'b}
G_0 |b^{(3)}\rangle
+\sum\limits_{c'(\subset s)}
{\sum\limits_{c(\subset c')}
{\left( {f_s^\dagger } \right)_{c'c}G_0|c^{(3)}\rangle \tau^{(3)}_c
\left( {x_s} \right)_{c'c,b'b}}}
\nonumber \\
&
+{\cal V}_sg_0\left( {x_s}^\dagger \right)_{-,b'b}
\label{s-first}
\\
\left( {x_s} \right)_{a'a,-}=&\langle a^{(3)}| G_0 \left( {f_s}
\right)_{a'a}
+\sum\limits_{c'(\subset s)} {\sum\limits_{c(\subset c')}
{\bar \delta _{ac}\delta _{a'c'}\langle a^{(3)}|G_0|c^{(3)}\rangle
\tau^{(3)}_c \left( {x_s} \right)_{c'c,-}}}\nonumber \\
&+\langle a^{(3)}| G_0 \left( {f_s} \right)_{a'a} g_0\left( {x_s} \right)_{-,-}
\label{s-last} \\
\left( {x_s} \right)_{-,-}=&{\cal V}_s+{\cal V}_sg_0
\left( {x_s} \right)_{-,-}
+\sum\limits_{c'(\subset s)} {\sum\limits_{c(\subset c')}
{\left( {f_s^\dagger } \right)_{c'c}G_0|c^{(3)}\rangle \tau^{(3)}_c
\left( {x_s} \right)_{c'c,-}}}
\label{last}
\end{eqnarray}
with $a\subset a'\subset s$ and $b\subset b'\subset s$.
For each $s$,
the subamplitudes $\left( {{\bf x}_s} \right)$
have components labelled by the Yakubovsk{\u\i} chain labels ($a'a$),
or by the symbol ``-" in the case of the no-pion sector, where the index
{\it s } specifies a unique physical partition (last column of table II in
\cite{Canton98}). Only when $s=0$, all couplings to the no-pion sector
are vanishing.
These coupled equations are obtained once
the representation Eq.~(\ref{apole}) has been assumed,
and the amplitudes are expressed  in the 4-body space in the
corresponding quasiparticle representation.
$G_0$ is the full 4-body free propagator,
$g_0$ is the already mentioned free three-nucleon propagator in the 
no-pion sector,
${\cal V}_s$ is the pair potential between the two interacting nucleons
in the partitions denoted by {\it s} and,
in addition to the short-range part of the 2N potential, it
includes explicitely the OPE diagram. Finally,
$f_s\;\left( {f_s^\dagger } \right)$
     is the elementary creation (annihilation) vertex for a pion into (from)
the appropriate subsystem denoted by the chain-label subscript.
An important aspect of these subamplitudes
is that it is always possible to factor out a Dirac $\delta$ function
in momentum space for the ``spectator" nucleon (or pion, for $s=0$).
The new aspect obtained in Ref.\cite{Canton98}
is that this factorization property has been maintained
when there is a pion associated with either the nucleon pair
or the spectator nucleon,  or when there is no pion at all.

The basic assumption for the $x_s$ subamplitudes
consists in the finite-rank representation
\begin{eqnarray}
\label{supersep1}
(x_s)_{a'a,b'b}&=& |(s^{(2)})_{a'a} \rangle \tau^{(2)}_s
                              \langle (s^{(2)})_{b'b}|\\
\label{supersep3}
({x^\dagger_s})_{-,b'b}&=&|(s^{(2)})_{-}\rangle \tau^{(2)}_s
                              \langle (s^{(2)})_{b'b}|\\
\label{supersep2}
(x_s)_{a'a,-}&=& |(s^{(2)})_{a'a}\rangle \tau^{(2)}_s
                              \langle (s^{(2)})_{-}|\\
(x_s)_{-,-} &=& |(s^{(2)})_{-}\rangle \tau^{(2)}_s
                              \langle (s^{(2)})_{-}| \,.
\label{supersep4}
\end{eqnarray}
This representation provides all the ingredients needed
to construct the connected dynamical equation (\ref{eLe}).
We have limited here the discussion to the case of
one separable term, but the algebraic generalization of Eq.(\ref{eLe})
to more separable terms is straightforward.

\section{Approximation scheme}

Although this is a very relevant issue,
we will not concentrate here on the mathematical aspects
and general constraints needed to obtain a mathematically converging
separable expansion for the subamplitudes ${\bf x}_s$.
A very clear explanation about the general methods required for
considering such questions can be found in Ref.~\cite{Weinberg64}.

We will instead concentrate on the development of an
approximate scheme for the separable representation of such subamplitudes.
The proposed approximation scheme is based on the physical
assumption that in standard nuclear physics the picture
of nucleons interacting via realistic nucleon-nucleon potentials,
with the pion degrees of freedom ``frozen out'',
provides an acceptable first-order description of the 
low-energy/low-momenta dynamics.
The effects of including explicit pionic degrees of freedom beyond that
picture should then be considered only as dynamical ``corrections''.

If we consider Eq.~(\ref{last}), we observe that the last term
accounts in fact for the one-pion dynamics in the 2N subsystem
($s\ne 0$), once the OPE diagram has been taken out (because it is already
included in ${\cal V}_s$).
In particular $({x}_s)_{-,-}$ represents the complete, elastic 2N amplitude
in presence of a spectator nucleon. Obviously, $({x}_s)_{-,-}$
should be obtained from the coupled set of
Eqs.~(\ref{first}-\ref{last}), but we will identify
instead $({x}_s)_{-,-}$ with the conventional 2N $t$-matrix,
derived by the solution of
the standard 2N Lippmann-Schwinger equation
(in the presence of a spectator nucleon) using as input
the phenomenological NN potential. In other words, we set to zero
the last term of Eq.~(\ref{last}) and use for
${\cal V}_s$ the conventional NN potential which
includes in an effective way the contributions
from the pion-2N dynamics.
Note that there will be a price to pay for this; namely,
all {\it disconnected} dispersive effects to the 3N dynamics,
originated in the 2N subsystems by the dynamical equations
(\ref{first}-\ref{last}), will be approximated to zero.
This is a consequence of the fact that the dynamical description
of the pion degrees of freedom implied by these  equations
has been replaced with an istantaneous, effective 2N potential.
Implicitely, this approximation is assumed in all potential
approaches to the 3N problem, but it has not really been tested.
The exception is in Ref.~\cite{Sauer92}, where these
dispersive effects of the 2N subsystem have been sized in the extreme
situation where the $\pi$N interaction is entirely
represented by its coupling through a forward propagating
$\Delta$-isobar. Interestingly, the $\Delta$-mediated,
disconnected dispersive effects in the 3N system turned out to be
not negligible.
It is clear that this problem should be investigated further;
nevertheless we will not do this here since our aim is to
follow  in this respect the standard
potential approach to the 3N problem,
where the 3N dispersive effects generated in the 2N subsystems
are completely ignored.

Once we have accepted that the 2N dynamics in the {\em elastic}
channel is described by means of a phenomenological
nucleon-nucleon potential,
our description can be closely compared to the
usual, potential-based, quantum-mechanical 3N approaches
since the dynamical input of the two approaches appears to be identical.
Then, if we consider Eq.~(\ref{supersep4}), this reduces to
the well-known, standard separable
representation of the two-nucleon $t$-matrix, which we express in
the polar form
\begin{equation}
\left( {x_s} \right)_{-,-}\approx \left| \tilde s^{(2)}\right\rangle
\tilde \tau _s^{(2)}\left\langle \tilde s^{(2)} \right|\, , \label{pole}
\end{equation}
and this already gives the approximations
$\left| {\left( {s^{(2)}} \right)_-} \right\rangle$ $\approx$
$\left| \tilde s^{(2)}\right\rangle$, and $\tau^{(2)}$ $\approx$
$\tilde \tau^{(2)}$ which are needed for the determination of the
driving term (first contribution) and for the kernel of Eq.~(\ref{eLe}).
It should be noted that the form factors
$\left| \tilde s^{(2)} \right\rangle $
for NN interactions are,
in this approximation, not distinct from the form factors
$\left| {a^{(3)}} \right\rangle $
of the quasiparticle approximation at the three-cluster level,
Eq.(\ref{apole}).
Both come, in fact, from the pole approximation of the
elastic two-nucleon $t$-matrix. The only difference is that the
$\left| \tilde s^{(2)} \right\rangle $ factor refers
only to the 2N $t$-matrices and is
expressed in the (NN)+N two-cluster space, with one Jacobi coordinate
removed already.
The form factor $\left| {a^{(3)}} \right\rangle $,
on the other hand, refers to all the two-body $t$-matrices in
the 4-body space, and this includes also the $\pi$N $t$-matrices
in addition to the NN ones.
Thus, a convenient feature emerges from the
approach herein discussed:
only one standard pole approximation
(or expansion, in the more general case of higher ranks)
has to be made for the two-nucleon interaction, to be used
for the 2N $t$-matrices in both 4-body and 3N spaces.

At this point one fact must be stressed: namely,
if we do not consider further contributions,
our description precisely collapses into the
quantum-mechanical approach to the 3N system in terms of
a 2N potential, since the resulting 3N equations have
{\em exactly} the AGS form.
This is because all couplings to $\pi$NNN state are made via the
components $\left| {\left( {s^{(2)}} \right)_{a'a} }\right\rangle$
which would be set to zero in this case.
Thus, these components, together with the $s=0$ partition,
are fundamental for an explicit treatment of the
pion degrees of freedom in the 3N system, and we get an approximation
of the $\left| {\left( {s^{(2)}} \right)_{a'a} }\right\rangle$ factors,
to the lowest order,  by considering in particular Eq.~(\ref{s-last}).
Assuming that the leading contributions to $\left(x_s\right)_{a'a,-}$
are dominated by the pole structures
(see Eq.~(\ref{pole})) of $\left(x_s\right)_{-,-}$
in the last term of Eq.~(\ref{s-last}),
we obtain
\begin{displaymath}
\left(x_s\right)_{a'a,-} \approx
\langle a^{(3)}|G_0 \left( {f_s} \right)_{a'a}g_0\left|
\tilde s^{(2)} \right\rangle
\tilde \tau _s^{(2)}\left\langle \tilde s^{(2)} \right|,
\end{displaymath}
and considering Eq.~(\ref{supersep3}),
\begin{equation}
\left| {\left( { s^{(2)}} \right)_{a'a}} \right\rangle \approx
\left| {\left( {\tilde s^{(2)}} \right)_{a'a}} \right\rangle \equiv
{\left\langle {a^{(3)}} \right|}
G_0\left( {f_s} \right)_{a'a}g_0\left| {\tilde s}^{(2)} \right\rangle\, .
\label{destro}
\end{equation}
This provides the form factors needed for the second term of
Eq.~(\ref{driv}), which in this approximation represents
the contribution entirely
responsible for the explicit treatment of the pion dynamics in
Eq.~(\ref{eLe}), beyond that effectively  contained in the static
2N potential.
The $\left\langle {a^{(3)}} \right|$
on the left of this is needed because the
quasiparticle approximation has already been made for the pair
interaction in the four-body space, by Eq.~(\ref{apole}).

Eq.~(\ref{driv}) also requires the  ``bra" vectors
$\left\langle \left(s^{(2)}\right)_- \right|$ ,
$\left\langle {\left( {\tilde s^{(2)}} \right)_{a'a}} \right|$.
We simply construct the adjoints, by taking the
corresponding ``bra" states of the underlying quasiparticle
approximation Eq.~(\ref{pole}) of the standard 2N
\mbox{t-matrix}. In other words
\begin{equation}
\left\langle \left(s^{(2)}\right)_- \right| \approx
\left\langle \tilde s^{(2)} \right| \, ,
\end{equation}
and
\begin{equation}
\left\langle \left(s^{(2)}\right)_{a'a} \right| \approx
    \left\langle \tilde s^{(2)} \right|g_0\left( {f_s^\dagger }
\right)_{a'a}G_0\left| {a^{(3)}} \right\rangle\, .
\label{sinistro}
\end{equation}

In the extended 3N equation, Eq.~(\ref{eLe}),
we identify three types of new contributions
which take into account, to the lowest order,
the explicit pion dynamics.
These contributions modify the driving term
$Z^{(2)}$ of the standard AGS equation in a very selective way.
The first contributions enlarge
the number of components with  respect to the three standard Faddeev
components, with the addition of the fourth component ($s=0$)
specifying the partition
when the pion is set apart from the three nucleons.
Such contributions provide the new couplings between
$s=0$ and $s'\ne 0$, as well as with $s\ne 0$ and $s'= 0$.
These link the 3N Faddeev components with the partition
consisting of a three-nucleon cluster and a separated pion ($s=0$).
A second type of terms enter in the diagonal part, $s = s' \ne 0$, of
$Z^{(2)}_{ss'}$ while traditionally these elements have been assumed
to be vanishing in the standard 3N theory:
hence, these terms should be considered important since they
compare to zero in the standard AGS equation.
Finally, the second term of Eq.(\ref{driv}) provides corrections
also for $0 \ne s \ne s' \ne 0$, that is for the off-diagonal elements
of the driving term. Instances of such terms are shown in the last
three contributions of Eq.~(1.2$a$). They all represent modifications
that have to be added to the standard 3N driving term,
identified with the first contribution in Eq.~(\ref{driv}).
Thus, the new contributions coming from the explicit pion degrees of
freedom, provide minimal, though important, modifications to the standard
AGS formulation of the three-nucleon problem.

\section{Discussion}

In the previous section we have shown that it is possible
to treat approximately the pion dynamics in the
3N equations by introducing the new form factor
\begin{equation}
\left| {\left( {\tilde s^{(2)}} \right)_{a'a}} \right\rangle \equiv
{\left\langle {a^{(3)}} \right|}
G_0\left( {f_s} \right)_{a'a}g_0\left| {\tilde s}^{(2)} \right\rangle\, .
\label{inelasticFF}
\end{equation}
The operator $\left( {f_s} \right)_{a'a}$ has been defined in
Ref.~\cite{Canton98} in terms of the renormalized elementary $\pi NN$
vertex by means of the inclusion prescription
\begin{equation}
\left( {f_s} \right)_{a'a}=\sum_{i=1}^{3} f_i
\bar\delta_{ia}\delta_{i,a\subset a'} \delta_{a'\subset s}\, .
\end{equation}
Here, the label ``$i$" represents the $\pi N$ pair interacting
via the vertex $f_i$.

\begin{figure}
\centerline{
\Spicture{5.5 in}{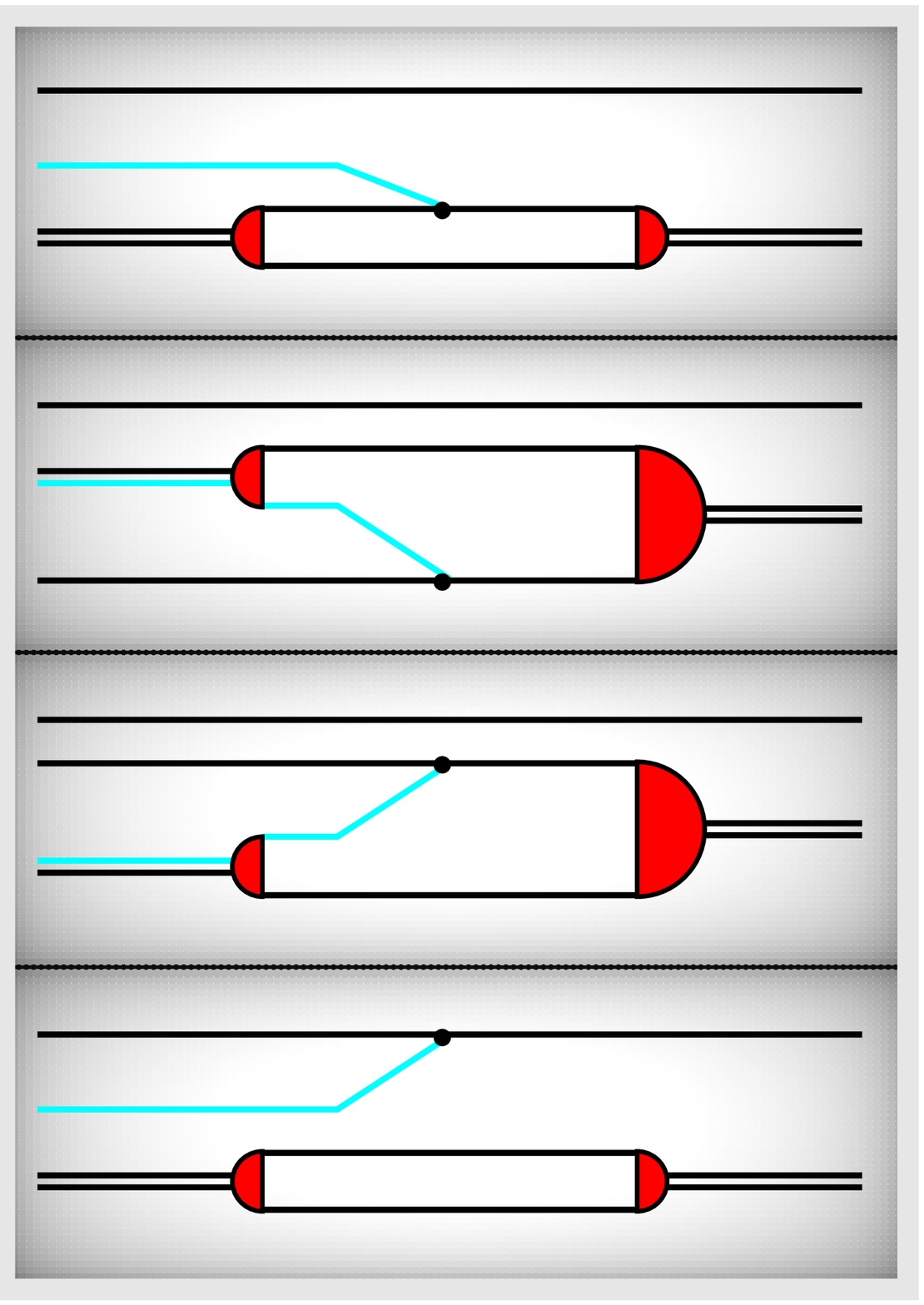}{}}
\vspace{3mm}
\caption{\label{diagramFF}
Diagrammatic representation of the form factors
$|\left(\tilde s^{(2)}\right)_{a'a}\rangle$
in the Yakubovsk{\u\i} chain-labelled space.
}
\end{figure}

In view of the key role played by this new ingredient in the extended 3N
equation we provide its detailed diagrammatic interpretation.
To fix the ideas, we choose $s=1$ and consider the various components
depending on the Yakubovsk{\u\i} chain-of-partition subscript $a'a$, listed
in Tab.~\ref{a'a}.
By applying the inclusion prescription for the vertex operator, we obtain
the set of diagrams drawn in Fig.~\ref{diagramFF}.
Note that there are only four diagrams
in the figure while there are five components in the table.
This is because the inclusion prescription automatically set to zero
the contribution corresponding to the last Yakubovsk{\u\i} component,
as shown in Tab~\ref{a'a}. Moreover, there is another diagram
(omitted in Fig.~\ref{diagramFF}) which has to be added to the first
diagram shown in the figure,
obtained by interchanging the two nucleons ``2" and ``3" within the
pair. This sum is evidenced in the first row of Tab.~\ref{a'a}.
All these contributions represent the proper quasiparticle
generalization of the standard 2N form factor,
in presence of a spectator nucleon, to the pion inelastic channel.

Once we have illustrated the form factor diagrams, we can discuss
the detailed structure of the new contributions to the driving term of the
extended 3N equation. Indeed,
we can interpret the main result obtained in Sect. III of
Ref.~\cite{Canton98}, and reported here in Eqs.(\ref{eLe},\ref{driv}),
as a simple prescription to include the pion dynamics
in the AGS equation,
\begin{equation}
Z_{ss'}=Z_{ss'}^{AGS}+Z_{ss'}^{\pi}
\end{equation}
and the main result of the previous section is a practical
approximation scheme
to get the part of the driving term which handles the pion dynamics
when $s, s' \ne 0$
\begin{equation}
Z_{ss'}^{\pi} =
\sum\limits_{a,a',b'}
\left\langle \tilde s^{(2)} \right|
g_0
\left(f^\dagger_s\right)_{a'a} G_0 \left|a^{(3)}\right\rangle
\tau _a^{(3)}
\left\langle a^{(3)}\right| G_0 \left(f_{s'}\right)_{b'a}
g_0
\left| \tilde s'^{(2)} \right\rangle
(\bar \delta _{ss'}+\delta _{ss'}\bar \delta _{a'b'}) .
\label{Z-pionic}
\end{equation}

\begin{figure}
\centerline{
\Spicture{5.5 in}{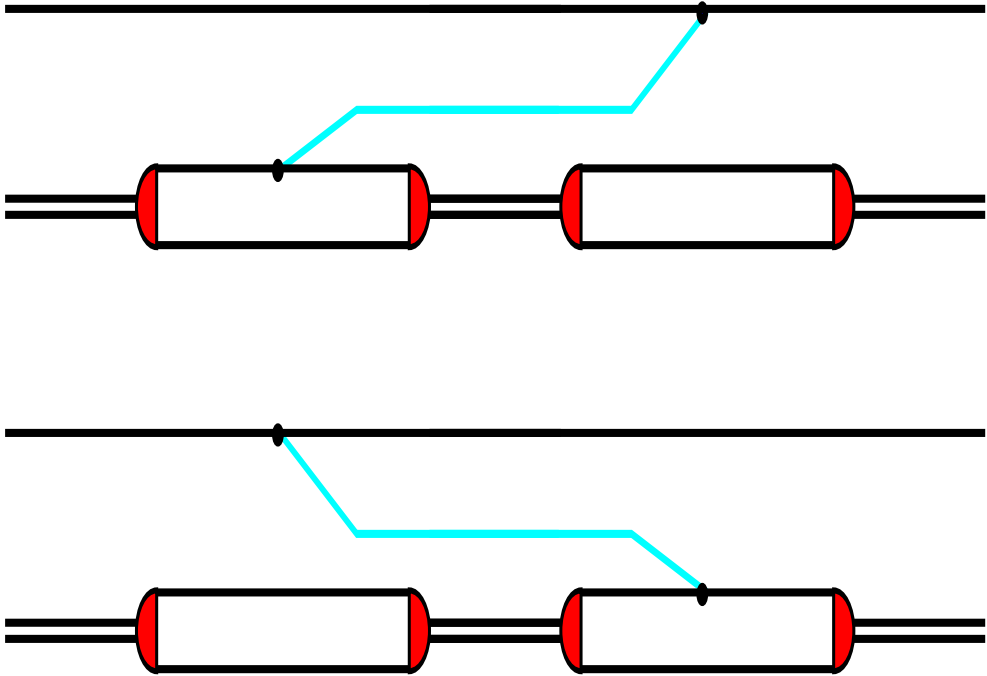}{}}
\vspace{4mm}
\caption{\label{s=s'} Diagrams contributing to $Z_{11}^{(2)}$.
In conventional 3N theory $Z_{11}^{(2)}$ is zero.
}
\end{figure}

We begin with the contributions to Eq.~(\ref{Z-pionic}) in the $s=s'=1$ case.
Here, only the term with $\delta_{ss'}$ survives in Eq.~(\ref{driv}),
which implies $a'\ne b'$, and also that only the ($N_2N_3$) pair
can be formed in the intermediate state. Consequently, the only possible
diagrams that can be constructed are obtained from
only the first and last diagrams in Fig.~\ref{diagramFF}
yielding the diagrams shown in Fig.~\ref{s=s'},
plus obviously those obtained by interchanging the pairing
nucleons ``2" and ``3", for a total of four time-ordered diagrams.
Clearly, these are irreducible contributions that cannot be
represented in conventional 2N potential theory.

\begin{figure}
\centerline{
\Spicture{5.5 in}{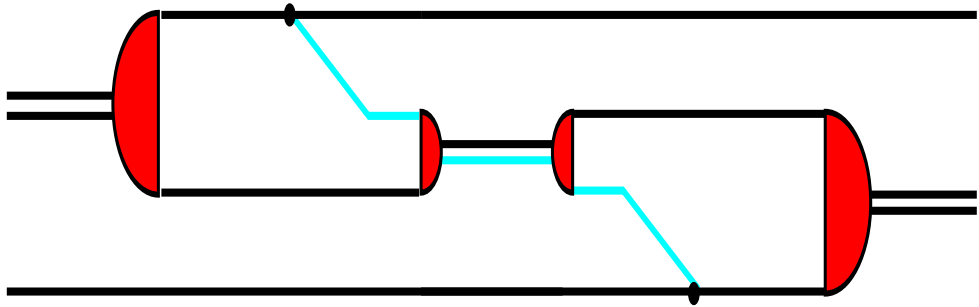}{}
}
\vspace{4mm}
\caption {\label{s_ne_s'}
The additional contribution to $Z_{12}^{(2)}$ due to
the treatment of the pion dynamics
}
\end{figure}

We then consider the contributions to the Z-pionic term
when $0\ne s\ne s'\ne 0$, {\it e.g.}, when $s=1$ and $s'=2$.
In this case only the  $\bar\delta_{ss'}$ contribution survives
in the second term of Eq.~(\ref{driv}); here the intermediate pair
can only be formed with the pion and the nucleon $N_3$,
which is uniquely defined since it does not act as spectator
in both components $s=1$ and $s'=2$. In this case,
one gets the diagram shown in Fig.~\ref{s_ne_s'}.
Note however that this is an approximated result:
Once we have approximated the inelastic form factors
in Eqs.~(\ref{supersep1}-\ref{supersep3}) by the leading expressions
(\ref{destro}, \ref{sinistro}),  then the last two terms in 
Eq.~(1.2$a$) vanish, because the last row in Tab.~\ref{a'a} 
is empty in this case.
However, in the more general approach of Ref.~\cite{Canton98}
there are additional contributions (shown in Fig.~3 of Ref.~\cite{Canton98})
which originate from the last two terms in Eq.~(1.2$a$).
Physically, these additional terms represent four-body
multiple-rescattering contributions.

%
%
%
So far, we have discussed the additional contributions one must
include in the AGS equation to take minimally into account
the pion dynamics beyond the OPE term. We observe that in both
cases these diagrams represent, in fact,
contributions that can be reinterpreted as
irreducible 3N potentials.  In particular, the
diagram of Fig.~\ref{s_ne_s'} has the
topological structure of the well-known
Fujita-Miyazawa\cite{fujmiy} diagram.  Similar
forms (where the exchanged pion rescatters before
being absorbed), are the basic ansatz for
building the pion part of the irreducible 3N
forces, such as the
Tucson-Melbourne\cite{tucmel},  Ruhr\cite{eden},
Brazil\cite{brasil}, or Texas\cite{texas} 3N interactions. However,
this scheme provides at the same time also
another set of irreducible 3N diagrams which must
be considered as well; these are  given in
Fig.~\ref{s=s'} and represent a totally different
structure from that of Fig.~\ref{s_ne_s'}. These
irreducible diagrams have been proposed by
Brueckner {\it et al.}\cite{BLM}, as early as
1954, and have been investigated
quantitatively for the triton binding energy by
Pask\cite{pask}, who finds they give a large
contribution. Since that time, it has been
argued  (see Refs.~\cite{eden,yang})
that these terms cancel out against relativistic
corrections to the  twice-iterated
one-pion exchange term.
However, in a recent study~\cite{Canton2000},
the effect of this cancellation has been questioned and in a
quantitative analysis the cancellation turned out to be remarkably
incomplete, with a breaking effect of 15-30\%.
The reason that breaks the cancellation is dynamical,
and can be evidenced with an approach preserving the
cluster sub-structures of the multinucleon system, while
describing the pion-exchange dynamics. This is, we
believe, one of the advandages of using the formulation
discussed here. From the diagrams of Fig.~\ref{s=s'}
a new contribution to the 3N force has been
extracted~\cite{Canton2000}, and by means of this contribution
the third nucleon affects in particular the triplet-odd waves
of the 2N subsystem, with possible consequences for the N-d vector
analyzing powers.

\begin{figure}

\centerline{
\Spicture{5.5 in}{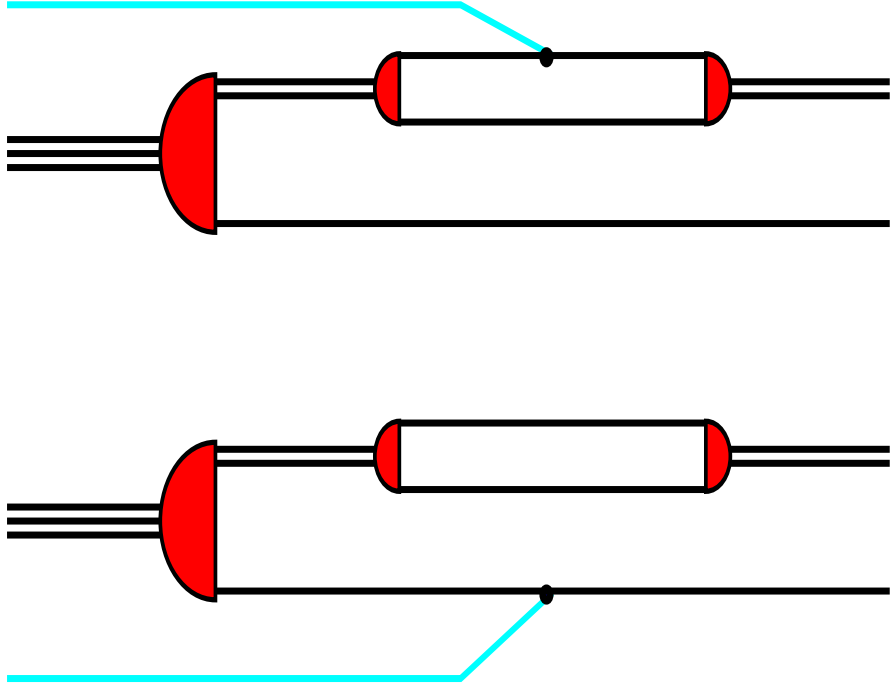}{}
}
\vspace{4mm}
\caption {\label{s=0_s'=1}
Processes contributing to $Z_{01}^{(2)}$ due to
the treatment of the pion dynamics
}
\end{figure}

Finally, we consider the contributions of the Z-pionic term
when $s=0$ and $s'\ne 0$. Here, the pionic contributions are crucial
in providing the couplings to the new, fourth component.
In such a case, one has to modify Eq.~(\ref{Z-pionic}) since the
approximation discussed in the previous section concerns the
structure of the $|s^{(2)}\rangle$ form factors only when
$s\ne 0$, while nothing is said otherwise. However, the
$s=0$ case has been discussed in Ref.~\cite{Canton98},
see Eqs.(2.16-18) therein, and it turns out that for $s=0$ the only
nonvanishing subamplitude is given by the equation
\begin{equation}
\left( {x_0} \right)_{a'a,a'b}=\langle a^{(3)} | G_0 | b^{(3)} \rangle
\bar \delta _{ab}
+ {\sum\limits_{c(\subset a')}
\langle a^{(3)} | G_0 | c^{(3)} \rangle
   {\bar \delta _{ac}\tau^{(3)}_c
\left( {x_0} \right)_{a'c,a'b}}} ,
\label{s=0case}
\end{equation}
which represents a Faddeev equation for three interacting
nucleons in presence of a spectator pion, since $a'=(NNN)\ \pi$.
In the vicinity of its poles, ${\bf x}_0$ becomes
\begin{equation}
(x_0)_{a'a,a'b}\simeq |(s^{(2)})_{a'a} \rangle \tau^{(2)}_s
                             \langle (s^{(2)})_{a'b}| \, .
\end{equation}
The $s=0$ form factors are solutions of the homogeneous
equation associated to Eq.~(\ref{s=0case}), and represent
the virtual decay of a 3N interacting cluster into a correlated
2N-pair, $a$, plus a nucleon, in presence of the spectator pion.
For $s=0$, the form factor 
is obviously vanishing in the pure 3N sector, since there
is always the presence of the spectator pion.
Hence, taking $s=0$ and $s'=1$ for instance, the Z-pionic term
becomes
\begin{equation}
Z_{ss'}^{\pi} =
\sum\limits_{a,a',b'}
\left\langle (s^{(2)})_{a'a} \right|
\tau _a^{(3)}
\left\langle a^{(3)}\right| G_0 \left(f_{s'}\right)_{b'a}
g_0
\left| \tilde s'^{(2)} \right\rangle,
\label{Z-pionic01}
\end{equation}
once the complicated $\delta$-structure in Eq.~(\ref{driv})
has been properly taken into account.
We observe that for $s\ne s'$ it is $a'\ne b'$ always.
Furthermore, since $a'$ identifies the $s=0$ partition,
$a$ may represent only a $NN$ pair, which means that
the term in Eq.~(\ref{Z-pionic01}) selects only
the first and the fourth Yakubovsk{\u\i} components (in Tab.~\ref{a'a}),
and this specifies completely the diagrams contributing
to $Z_{01}^{\pi}$,  as shown in Fig.~\ref{s=0_s'=1}.
In the figure the first diagram should include an additional contribution
obtained by interchanging the pairing nucleons, as usual.

At this point, we have achieved
the main goal of this work; indeed,
starting from the more general approach of Ref.~\cite{Canton98},
we derived a practical,
approximated scheme for the treatment of the pion dynamics in
the 3N system. In particular, we obtained a set of dynamical equations
where the main input is given by the  2N $t$-matrix
generated by phenomenological NN potentials (hence,
constrained by phase-shift analysis).
The additional inputs required for this description
are the $\pi N$ $t$-matrix, once its polar part has been subtracted,
and the nonrelativistic $\pi$NN vertex, needed for the construction of
new form factors expressed by Eq.~(\ref{inelasticFF}).

This set of equations can be considered a
natural, approximated extension of 3N AGS equations for
the explicit treatment of the pionic channel.
We have identified the modifications implied by this treatment:
they consist in additional terms, $Z_{ss'}^\pi$,
representing corrections to the standard
AGS driving term $Z_{ss'}^{AGS}$. We have classified
these corrections by the structure of the underlying diagrams
and found they all correspond to irreducible 3N-force diagrams.
We have also pointed out the presence of a fourth, pionic, component
representing the $\pi$ + (NNN) partition, and described how this is
coupled to the other three Faddeev components.
Finally, we have discussed the main limitation implied
by the approach, which consists in ignoring all 3N disconnected
(or reducible) dispersive effects: this is unavoidable if we use
as input a potential-based description for the dynamics of the 2N
subsystem;
indeed, such an approximation is implicitly assumed in all
conventional quantum-mechanical descriptions of the 3N system based on
the potential approach. To overcome this limitation,
one should consider the more general approach of
Refs.~\cite{trabam,Canton98}, wherein the input interactions cannot
be defined in a simple manner.
With respect to this point, we observe that, since
the 2N-subsystem dynamics is described in this work
in terms of the standard  2N potential approach, the mass of the
nucleon in the no-pion sector is not generated dynamically,
but is static, and hence these approximated equations are not
plagued by the nucleon-renormalization problem\cite{Sauer85}.
On the other hand, above the pion threshold for the NN subsystem,
the approach presented herein can not be considered unitary,
since the input 2N $t$-matrices are not unitary either.

It might be surprising to find out that the dynamical
equations analyzed here as well as in Ref.\cite{Canton98}
lead to another, new class of irreducible 3N diagrams.
  Such a class of diagrams is present but still hidden
  in these dynamical equations since these represent
  formulations suited for the 3N problem above the pion threshold.
  If we consider the 3N system at lower energies, then it is more convenient
  to eliminate from these formulations the fourth Faddeev component,
  $s=0$, related to the mesonic channel. This can be accomplished
  by using in a very straightforward way Feshbach's projection technique.
  The resulting dynamical equations have only
  three Faddeev components, just like the standard AGS formalism,
  while the mesonic channel has been projected out from
  Eq.~(\ref{eLe}). This recasting of the dynamics
  (which represents an exact result)
  leads to an additional modification of the driving term $Z_{ss'}$,
  due to  the effects of the coupling to the fourth component,
  \begin{equation}
  Z_{ss'}=Z_{ss'}^{AGS}+Z_{ss'}^{\pi} + Z_{ss'}^{'} \, .
  \label{Z-AGS-extended}
  \end{equation}
  (Here and in the following, it must be assumed that
  the indices $s$, and $s'$ span only from 1 to 3,
  since the mesonic channel has been projected out.)
  The additional term, $Z_{ss'}^{'}$, is given by
  \begin{equation}
  Z_{ss'}^{'}=Z_{s0}^{\pi}\tau^{(2)}_0 Z_{0s'}^{\pi}\, ,
  \end{equation}
  while $Z_{00}^\pi$ is identically zero
  as has been pointed out in the introduction.

\begin{figure}

\centerline{
\Spicture{5.5 in}{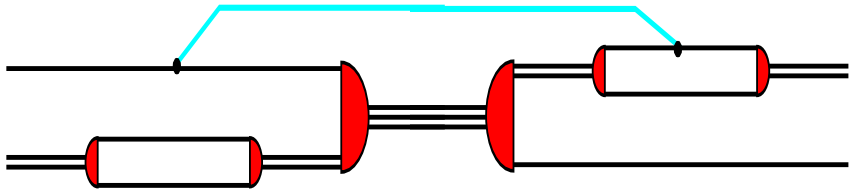}{}
}
\vspace{4mm}
\caption {\label{newclass}
Intermediate 3N cluster formation while the meson is in flight.
Processes contributing to $Z_{ss'}^{'}$
and generated by (projecting out) the fourth Faddeev component.
}
\end{figure}

  We consider this new
  additional contribution to the driving term, $Z_{ss'}^{'}$,
  and discuss what kind of irreducible 3N diagrams are involved.
  This can be accomplished by considering the definition of $Z_{0s}^\pi$
  given in Eq.~(\ref{Z-pionic01}), and taking into account the
  fact that $\tau_0^{(2)}$ represents the strength of the connected
  3N correlations while the pion is ``in flight", as discussed
  in Ref.~\cite{Canton98}.
  A strong 3N correlation would correspond to a pole-like structure
  for $\tau_0^{(2)}$, shifted by the energy carried by the exchanged pion.
  Using the result expressed in Eq.~(\ref{Z-pionic01})
  one can show that the term $Z_{ss'}^{'}$
  can be written as
  \begin{eqnarray}
Z_{ss'}^{'} =&
\sum\limits_{b,c,b',c'}
\left\langle \tilde s^{(2)} \right|
g_0 \left(f^\dagger_{s}\right)_{b'b}
G_0 \left| b^{(3)}\right\rangle
\tau _b^{(3)}
\left| (0^{(2)})_{a'b} \right\rangle
\tau_0^{(2)} \\ \nonumber
&\times
\left\langle (0^{(2)})_{a'c} \right|
\tau _c^{(3)}
\left\langle c^{(3)}\right| G_0 \left(f_{s'}\right)_{c'c}
g_0
\left| \tilde s'^{(2)} \right\rangle,
\label{Z^(0)}
  \end{eqnarray}
and  the diagram shown in Fig.~\ref{newclass}
is just one contribution to this structure. Note that the pion
can couple any of the three ingoing nucleon lines
with each one of the three outgoing lines. Moreover
this occurs with all possible recombinations
in intermediate 2N pairs.

This set of diagrams represent another, new class
of irreducible 3N mechanisms contributing to the
construction of the 3N force. Physically, these diagrams
represent all possible {\it connected} correlations
among the three nucleons while the exchanged pion is
in flight.  On the contrary,
the diagrams of Fig.~\ref{s=s'} represent all possible
{\it disconnected} 2N correlations  while the pion is in flight.

\section{Summary and Conclusions}

Recently, a new approach for the explicit treatment of the pion dynamics
in the 3N system has been obtained~\cite{Canton98}. Herein, we have shown
how to derive from this formulation a practical calculation scheme
which is phenomenologically sound.
The approximation is based on the assumption that the elastic
2N subamplitudes can be conveniently described with the phenomenological
2N potential approach. In other words, instead of treating explicitly
the pion dynamics in full, in the 3N system we describe explicitely
only those aspects of the pion dynamics which cannot be
buried into the all-comprehensive, phenomenological 2N potential.
Thus, the procedure could be viewed as a method to cool down
(or to gradually project out) the pion dynamics from the theory of
Ref.~\cite{Canton98}.

One advantage of making such an approximation
is that this approach is not plagued with the nucleon-renormalization
problem, since the dynamical equation used to construct the subsystem
amplitudes is represented by the standard 2N Lippmann-Schwinger equation,
where the nucleon masses are static. However the approach includes
also the pion degrees of freedom, via the inelastic subamplitudes.
These are represented as 2N form factors defined in the
Yakubovsk{\u\i} chain-labelled space of the (four-body)
$\pi$NNN system. Here, we consider only the first-order
contributions to such inelastic form factors in terms of the
effective $\pi$NN coupling vertex, see Tab.\ref{a'a}.
(It is evident that more complex
rescattering mechanisms contributing to such form factors can be
implemented at a later stage.)
Another advantage of this approach is that the dynamical input can be
constrained by the 2N experimental data;
therefore the method can be directly compared
with the standard quantum-mechanical approach to the 3N problem
based upon a 2N potential description.

By using a separable-expansion representation of the elastic NN $t$-matrix,
it has been possible to recast the dynamical equation into an extended
AGS form, where the part of the pion dynamics not buried
in the 2N potential is treated explicitly. We discussed the
three modifications
implied by this extended 3N equation with respect to
the normal AGS one.

First, the 3$\times$3 AGS driving term, defined in
terms of the three Faddeev components, acquires additional
contributions from the pion dynamics, and these new
contributions act in both diagonal and off-diagonal matrix elements
(while the standard AGS driving term
is known to act
only in the off-diagonal elements). We discussed these contributions
also in terms of their diagrammatic interpretation, and found that
the off-diagonal corrections correspond, to their lowest order,
to irreducible
3N-force diagrams where the pion, while being exchanged between two
nucleons, rescatters from the third before annihilating.
This rescattering mechanism is practically the only case discussed
in the construction of the pionic part of the 3N force.
And the different 3NF approaches differ mainly for the model representation
of the $\pi$N rescattering amplitude; there are,
of course, additional
short-range effects where there is much more ambiguity
and where the various 3NF models differ considerably among each other.
Conversely, the second modification refers to the diagonal
part of the driving term, and  represents
irreducible 3N-force diagrams of different topology, where
it is one of the two nucleons exchanging the pion that
rescatters with the third one.
These 3NF diagrams are not included in the construction of
modern 3N potentials because it is usually assumed that
they cancel out if one takes into account meson retardation effects
in all their possible variety of time orderings.
However, as has been pointed out in a
recent study~\cite{Canton2000}, this cancellation
as a 100\% effect is questionable.
The reason for questioning the cancellation is due to the fact
that a full 2N sub-amplitude enters in this 3N-force diagram,
while the cancellation involves only the ``instantaneous"
part of the diagram. The construction of 3NF models
from meson-exchange mechanisms evaluated in terms of instantaneous
processes will lead inevitably to a 100\% cancellation effect;
however such methods do not take into account that two nucleons may
clusterize while the pion is being exchanged, and this implies
that an entire set of rescattering processes have to be subsummed
during the pion-exchange process, thus leading to an incomplete
cancellation effect.

The third and  last modification with respect to
the normal AGS equation implies the
increasing of the matrix dimension by one unit, since the 3 Faddeev
components are now coupled to the additional two-cluster partition
$\pi$+(NNN). We have discussed the structure of such couplings
and the corresponding diagrams, thus providing details for all
the ingredients of this new dynamical equation.

Finally, we have shown that it is possible to project out the effects
of the coupling to this pionic channel, thus providing
a 3N dynamics with explicit treatment only for the 3N coordinates.
This modification
is suited for treatments of the 3N system at lower
energies, below the pion threshold. Then, the extended AGS equation
involves only the three standard Faddeev components, and the effect
of the $\pi$+(NNN) channel in the intermediate states
is contained into an additional, third contribution to the driving
term. We have analyzed the diagrams involved and found that
they correspond to a third class of irreducible 3NF diagrams,
topologically different from the other, previously discussed
two classes. The diagrams correspond to the variety of
connected  correlations amongst the three nucleons while the
meson is in flight. In other words, these mechanisms
take into account all possible
3N-cluster effects while the meson is being exchanged. This
new correction acts in both diagonal and off-diagonal matrix elements
of the driving term and may possibly influence the structure
of the 3N force.

\acknowledgements

L.C. acknowledges funds from the Italian Murst-PRIN Project
``Fisica Teorica del Nucleo e dei Sistemi a pi\'u corpi",
and hospitality and financial support from the University of Manitoba,
during two visits in 1999.
J.P.S. acknowledges continuing financial support from NSERC, Canada.
T.M. is grateful to the University of Manitoba for scholarship support.
J.P.S. and T.M. are thankful to INFN, Padova, for hospitality and financial
support during recent visits to Padova.

\newpage


\begin{table}

\begin{center}

\begin{tabular}{|c|c|c|}
\hline
$a'$ & $a$ & $(f_1)_{a'a}$\\
\hline
$(N_2\ N_3\ \pi) \ N_1 $ & $(N_2\ N_3)\ \pi\ N_1$& $f_2+f_3$  \\
$(N_2\ N_3\ \pi) \ N_1 $ & $(N_2\ \pi)\ N_3\ N_1$&   $f_3$ \\
$(N_2\ N_3\ \pi) \ N_1 $ & $(N_3\ \pi)\ N_2\ N_1$&   $f_2$ \\
$(N_2\ N_3)\ (\pi\ N_1)$ & $(N_2\ N_3)\ \pi\ N_1$&   $f_1$ \\
$(N_2\ N_3)\ (\pi\ N_1)$ & $N_2\ N_3\ (\pi\ N_1)$&    $-$  \\
\hline
\end{tabular}
\end{center}
\caption{\label{a'a}
Chain-labelled 4-body components of the form factor
$|\left(\tilde s^{(2)}\right)_{a'a}\rangle$ for $s=1$.
The last column represents how this form factor is constructed
in terms of the elementary $\pi$ interaction
with nucleon $i$, by means of the $\pi$NN vertex $f_i$}
\end{table}

\end{document}